# PARADOXES OF THERMAL EXPANSION

## I.A. STEPANOV[1]

*Institute of Physical Chemistry II, Albertstrasse 23a, Freiburg 79104, Germany*

It is shown that the dependence of negative thermal expansion coefficient of many substances on the temperature contradicts to an important thermodynamic relation $(\partial C_P/\partial P)_T = -TV((\partial\alpha/\partial T)_P + \alpha^2)$. It is supposed that there are oscillations at the $C_P(T)$ and $\alpha(T)$ curves at $\alpha>0$ and they are in reverse phases.

In [1-11] dependence of the thermal expansion coefficient $\alpha$ and heat capacity $C_P$ on the temperature has been obtained for a number of substances. There is a thermodynamic relation using which one can verify the 1st law of thermodynamics [12, 13]:

$$(\partial C_P/\partial P)_T = -TV((\partial\alpha/\partial T)_P + \alpha^2) \qquad (1)$$

One can show that the dependence $\alpha(T)$ found in [1-11] contradicts to the relation (1) for negative $\alpha$. In al temperature intervals under the scope of this paper, $\alpha^2 << |(\partial\alpha/\partial T)_P|$ and can be omitted. In [14, 15] this contradiction has been shown for $ZrW_2O_8$ and $HfW_2O_8$.

In [1-5] $\alpha(T)$ and $C_P(T)$ have been obtained for $CeAl_3$: Figures 1-5. From them one can deduce that $(\partial\alpha/\partial T)_P<0$ at $0<T<0.4$ K. However, $(\partial C_P/\partial P)_T<0$ in this interval. Both sides of (1) have different signs.

In all references where dependence $C_P(P)$ is given, $\text{sign}(\partial C_P/\partial P)_T = \text{sign}(dC_P/dP)$. $C_P=C_P(T, P)$. A special case is $C_P=C_P(T(P))$:

$$dC_P/dP = (dC_P/dT)dT/dP \qquad (2)$$

$dT/dP<0$ for $\alpha<0$.

In the biggest part of the phenomena of nature and technique $P=P(T)$. (For example, it is necessary to calculate pressure in a vessel caused by negative thermal expansion of substance).

For $\alpha<0$, $dT/dP<0$ and $dC_P/dT>0$ whence the left part of (1) $(\partial C_P/\partial P)_T<0$. In [3] dependence $\alpha(P)$ has been found also in the pressure range 0.4-8.2 kbar. In this range $CeAl_3$ has no negative expansion.

At $T>0.4$, $(\partial\alpha/\partial T)_P>0$ and $(\partial C_P/\partial P)_T>0$. Again both parts of (1) have different signs.

---

[1] *Permanent address: Institute of Chemical Physics, Latvian University, Riga, Rainis bulv. 19, LV-1586, Latvia*

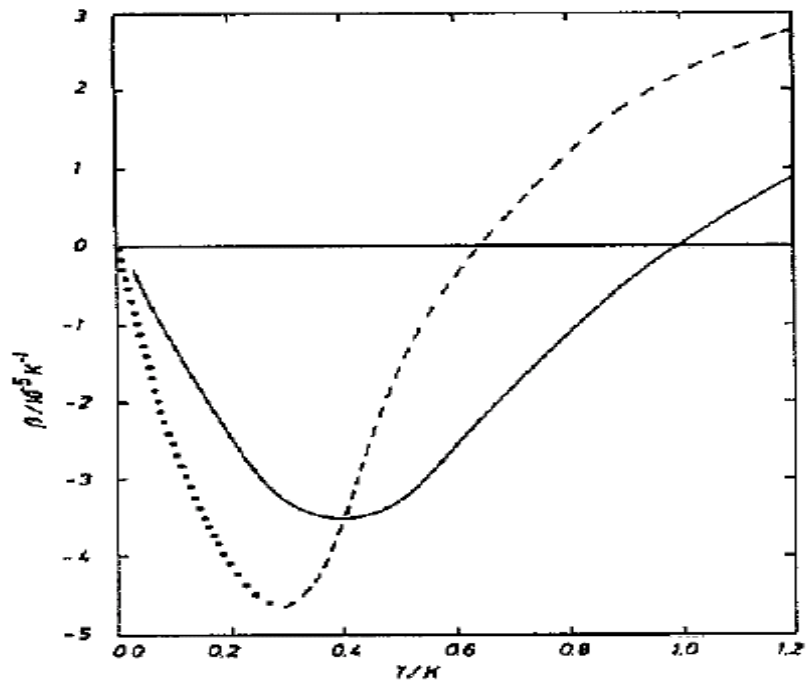

Figure 1. Coefficient of volumetric expansion for polycrystalline $CeAl_3$. , ———, [1], – – – –, [2], ········, [3] (extrapolation).

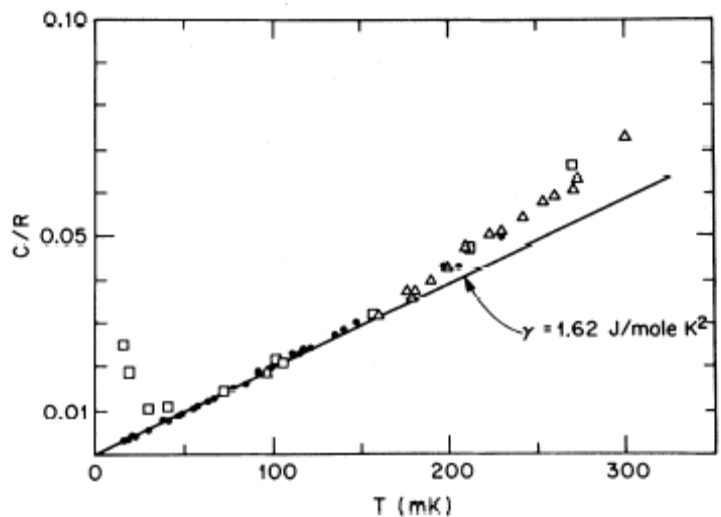

Figure 2. Specific heat of CeAl$_3$ at very low temperatures in zero field (●, ∆) and in 10 kOe (□) [2].

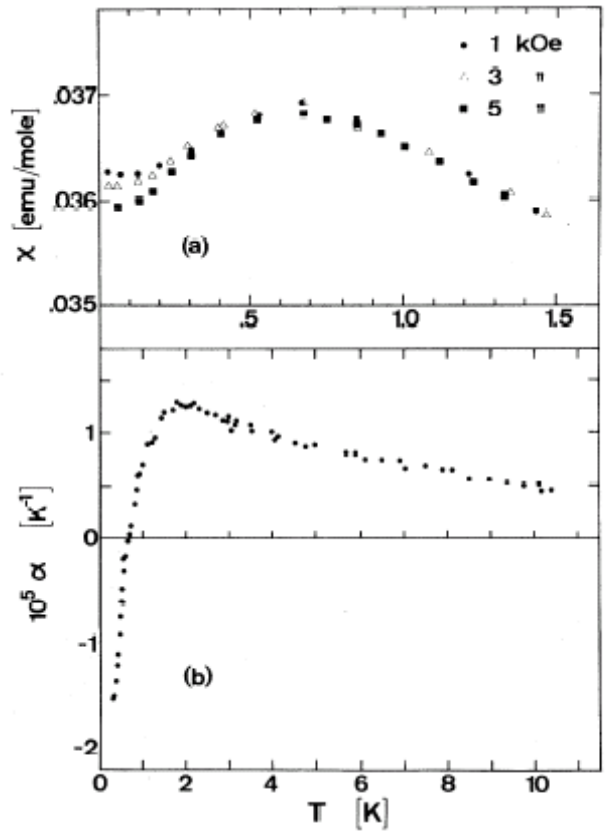

Figure 3. (a) Susceptibility of polycrystaline CeAl$_3$ in different magnetic fields below 1.5 K. (b) Linear thermal expansion coefficient of a polycrystaline sample of CeAl$_3$ below 10 K [2].

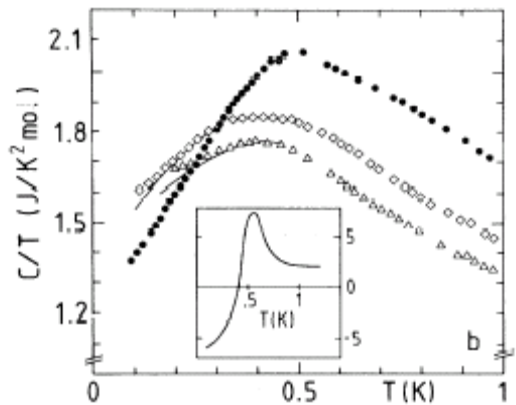

Figure 4. C/T vs T for CeAl$_3$, as a function of magnetic field (●, 0 Tesla, □, 1 Tesla, , 4 Tesla). The maximum experimental error is indicated by the symbol size. The inset shows $\partial \ln(C/T)/\partial \ln V$ vs T [4].

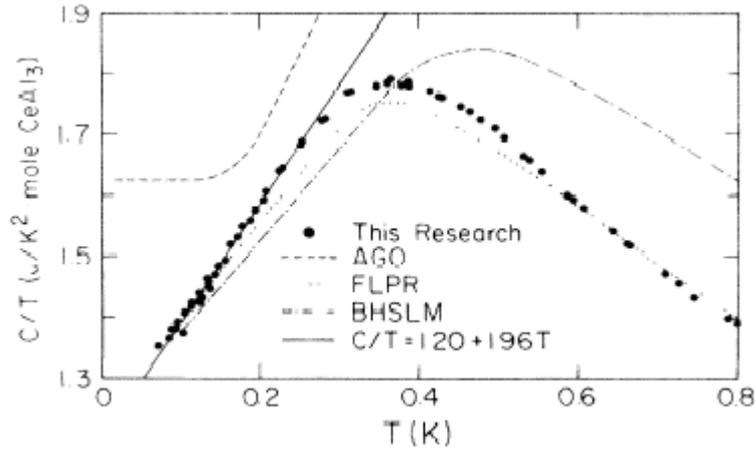

Figure 5. Specific heat of $CeAl_3$ for $T \leq 0.8$ K and $P=0$. Curves representing data from three other laboratories are also shown [5].

In [6] dependences $\alpha(T)$ and $C_P(T)$ were given for Si (Figures 6 and 7). It is clear that the data contradict to (1) at $0<T<85$ K. It is possible that the curve $C_P(T)$ decreases from $T \approx 85$ K till $T \approx 120$ K (the point of intersection of $\alpha(T)$ with the line $\alpha=0$) but the experimenters omitted the points. From $T \approx 120$ K it begins again to increase.

At $T>120$ K, $\alpha>0$, $(\partial \alpha/\partial T)_P>0$ and $(dC_P/dT)dT/dP>0$. There is again a contradiction to (1). One can suppose that there are oscillations at the $\alpha(T)$ and $C_P(T)$ curves and they are in reverse phases. One can detect them if to improve resolution. In [7] a peak was detected at the $C_P(T)$ curve at $T \approx 4$ K when resolution was improved. There dependencies $C_P(T)$ and $V(T)$ are given for UNiGe. There are positive peaks at the $C_P(T)$ curve and negative peaks at the $\alpha(T)$ curve at the same T. It contradicts to (1). The same situation is in [8] and there one can see large maximum and minimum (at the same temperature) at the $\alpha(T)$ and $C_P(T)$ curves of UPt respectively and they are in reverse phases.

In [8, 9] it is shown that negative expansivity of supercooled water gets more and more negative as temperature approaches 228 K starting from 273 K. At the same time, $C_P(T)$ increases. This behaviour contradicts to (1).

A possible explanation of these phenomena can be found in [13]. There it has been supposed that for substances with negative thermal expansion the 1st law of thermodynamics must have the following form: $dQ=dU-PdV$. If to derive (1) using this formula, one obtains

$$(\partial C_P/\partial P)_T = TV((\partial \alpha/\partial T)_P + \alpha^2). \qquad (3)$$

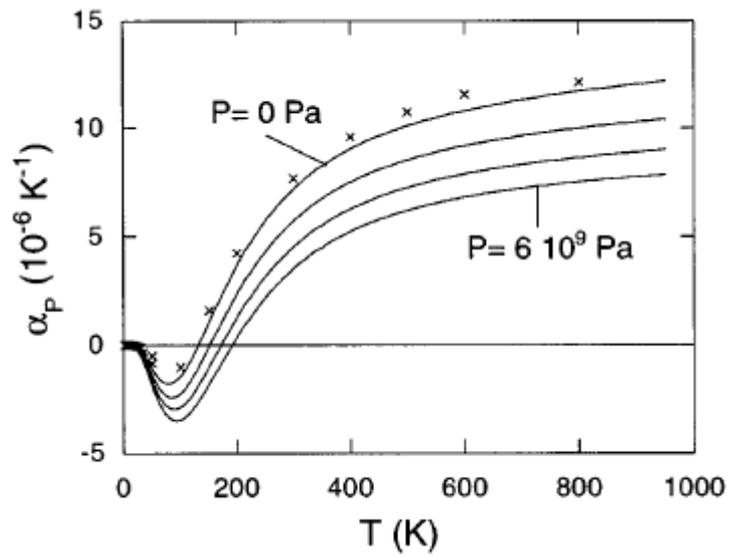

Figure 6. Temperature dependence of the volumic thermal expansion coefficient $\alpha_P$ of Si for four different pressures. The crosses indicate experimental data [6].

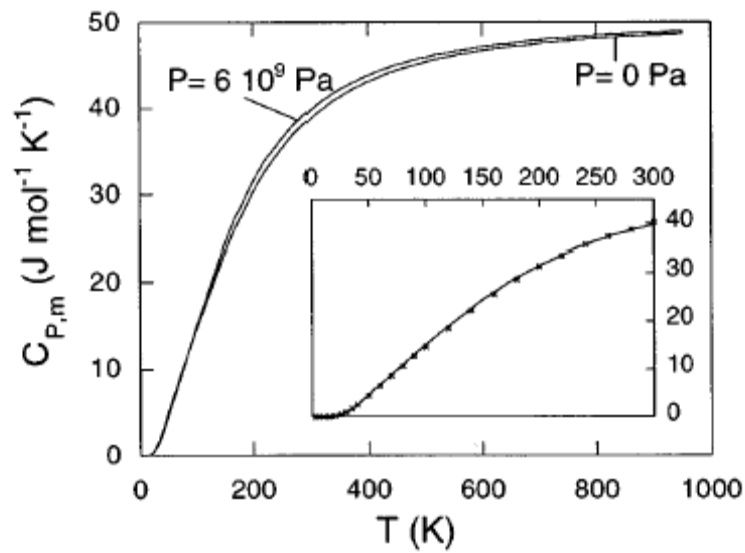

Figure 7. Temperature dependence of the constant-pressure specific heat $C_{P,m}$ of Si for two different pressures (calculation). There is a misprint on this Figure: 0 Pa is the upper curve at high temperature, $6 \cdot 10^9$ Pa is the lower curve at high temperature. The crosses indicate experimental data [6].